# Quench-Induced Degradation of the Quality Factor in Superconducting Resonators

M. Checchin[*] and M. Martinello

*Fermi National Accelerator Laboratory, Batavia, Illinois 60510, USA*
*and Department of Physics, Illinois Institute of Technology, Chicago, Illinois 60616, USA*

A. Romanenko, A. Grassellino, D. A. Sergatskov, S. Posen, and O. Melnychuk

*Fermi National Accelerator Laboratory, Batavia, Illinois 60510, USA*

J. F. Zasadzinski

*Department of Physics, Illinois Institute of Technology, Chicago, Illinois 60616, USA*

Quench of superconducting radio-frequency cavities frequently leads to the lowered quality factor $Q_0$, which had been attributed to the additional trapped magnetic flux. Here we demonstrate that the origin of this magnetic flux is purely extrinsic to the cavity by showing no extra dissipation (unchanged $Q_0$) after quenching in zero magnetic field, which allows us to rule out intrinsic mechanisms of flux trapping such as generation of thermal currents or trapping of the rf field. We also show the clear relation of dissipation introduced by quenching to the orientation of the applied magnetic field and the possibility to fully recover the quality factor by requenching in the compensated field. We discover that for larger values of the ambient field, the $Q$-factor degradation may become irreversible by this technique, likely due to the outward flux migration beyond the normal zone opening during quench. Our findings are of special practical importance for accelerators based on low- and medium-$\beta$ accelerating structures residing close to focusing magnets, as well as for all high-$Q$ cavity-based accelerators.

## I. INTRODUCTION

Superconducting radio-frequency (SRF) cavities are resonant structures that allow accelerating charged particles up to energies of tera-electron-volts [1–3]. The limiting factors of such accelerating structures are represented by the finite value of the intrinsic quality factor $Q_0$, directly related to the cryogenic cost needed for their operation, and by the radio-frequency- (rf-)field breakdown due to quench that limits the maximum achievable accelerating gradient $E_{\mathrm{acc}}$.

A typical quench event is initiated by a small area of the cavity surface becoming normal-conducting either due to heating up above the critical temperature ($T_c$) or due to the local critical field being exceeded. The sharp increase of the surface resistance in the normal zone can be contained only up to a certain dissipation level, above which a fast avalanchelike spreading of the normal zone that consumes all of the rf field in the cavity occurs. Several known mechanisms [1,2,4–6] may cause quench, and it was hypothesized that when the normal-conducting region is created, some magnetic flux can be trapped at the quench spot causing extra dissipation [7].

The origin of such trapped magnetic flux remained unclear and was ascribed to different mechanisms, such as thermal currents driven by the local thermal gradient in the quench zone [7], rf field trapped within the penetration depth region, or ambient magnetic field. However, a full understanding of the phenomenon has not been developed yet.

Previous studies [8–11] of the quality factor degradation in high- and medium-$\beta$ superconducting resonators targeted a criterion for the amount of flux trapped during the quench. A clear dependence of the quench-related degradation on the locally applied nonuniform external magnetic field was found, highlighting the possibility that extra dissipation introduced by quenching was of environmental origin. The "quench annealing" phenomenon—the recovery of the cavity quality factor by quenching when the additional field was removed—was also documented in these studies. Some deficiencies of these studies were the lack of advanced thermometry mapping and nonuniform magnetic field environment, making a full description challenging. Also, higher than typical magnetic field values were explored, similar to those found in special cases such as cavities operating close to a strong magnet.

In the presented study, we use the full range of current state-of-the-art techniques including advanced temperature mapping and Helmholtz coils to understand the detailed physics behind the quality factor degradation due to quench in superconducting resonators. In SRF applications, our results can be helpful for $Q$ preservation in accelerators utilizing cavities at very high-$Q$ values (requiring very challenging magnetic field shielding and cooldown process), as well as for designing cryomodules where SRF structures need to operate nearby sources of high magnetic field (usually solenoids or quadrupole magnets).

[*]checchin@fnal.gov

We report the experimental proof that the $Q_0$ degradation due to quench is a direct consequence of trapped *ambient* magnetic field, ruling out any other possible mechanisms. We also demonstrate that a full recovery of $Q_0$ after quench can be achieved when the cavity is quenched in the absence of the external magnetic field—an alternative to warming up above the critical temperature—and present a consistent physical model of this phenomenon. In addition, we find a dependence of the extra losses after quench on the orientation of the external magnetic field with respect to the cavity axis. To understand the recovery of $Q_0$, the key is the configuration of the magnetic field trapped at the quench spot, which we discuss in detail. We observe that the recovery of the quality factor is not possible if the externally applied field is big enough (>1 Oe). The proposed explanation for this irreversibility is the migration of the flux farther from the quench spot.

## II. EXPERIMENTAL SETUP

Quench experiments are performed using multiple 1.3-GHz fine-grain bulk Nb cavities of elliptical TESLA shape [12]. Three bare one-cell cavities and one dressed nine-cell cavity are prepared by nitrogen-doping recipes, and one one-cell cavity is prepared by a standard EP + 120 °C baking international linear collider (ILC)-type recipe (Table I). Nitrogen-doped cavities (nine-cell cavity included) are baked for 3 h at 800 °C before the doping treatment. All measurements are done at the FNAL cavity vertical test facility.

Schematics of the cavity instrumentation used are presented in Fig. 1.

In order to map the temperature variation over the cavity wall during quench, localize the quench spot site, and study in detail the resulting dissipation pattern, the one-cell cavities are equipped with the advanced temperature-mapping system ($T$ map) [14] based on an array of carbon-resistive sensors placed on a total of 36 boards— 16 per board—with the boards positioned every 10° around the cavity circumference. The external magnetic field is sustained by Helmholtz coils and measured by four single-axis Bartington Mag-01H cryogenic fluxgate magnetometers positioned at the equator axially to the cavity and spaced about 90° between each other; see Fig. 1(a) for the schematic. For one of the cavities (AES011), two sets of Helmholtz coils are used generating fields in two different directions (axial and orthogonal). In this configuration, no temperature mapping is used due to space constraints.

As shown in Fig. 1(b), the fully dressed LCLS-II nine-cell cavity (AES024) is equipped with two sets of Helmholtz coils and three fluxgate magnetometers placed outside of the helium vessel.

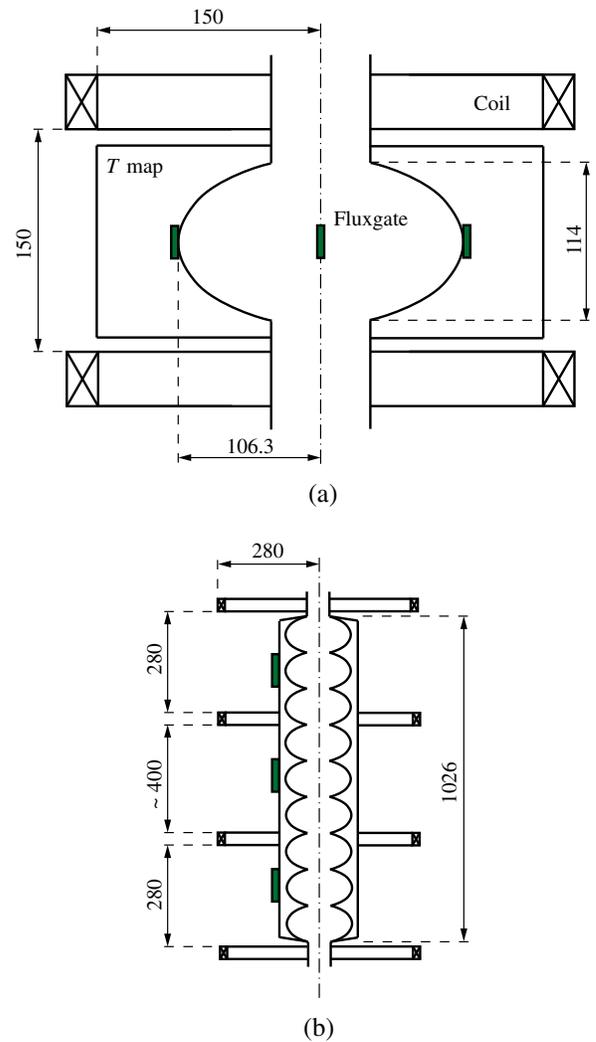

FIG. 1. Experimental setup for (a) one-cell cavities and (b) a nine-cell fully dressed LCLS-II cavity. All the dimensions are given in millimeters.

TABLE I. Cavities studied with respective thermal treatments and quench fields. Doped cavities are treated with 25 mTorr of $N_2$ and with a post-treatment chemistry (EP) of 5 $\mu$m.

| Cavity | Processing treatment | Cavity type |
|---|---|---|
| AES011 | 800 °C, 2 min with $N_2$ + 6 min without $N_2$ | Bare one cell |
| AES019 | 800 °C, 10 min with $N_2$ | Bare one cell |
| ACC002 | 800 °C, 20 min with $N_2$ | Bare one cell |
| AES014 | 120 °C bake | Bare one cell |
| AES024 | 800 °C, LCLS-II N-doping treatment [13] | Dressed nine cell |

In order to minimize the temperature-dependent part of the surface resistance, all the measurements except for the nine-cell cavity (measured only at 2 K) are done at the lowest temperature achievable by the cryoplant, which is around 1.5 K.

## III. RESULTS

All the measurements are performed by quenching cavities in the presence of the external magnetic field ($H$) or in compensated magnetic field and by recording the degradation of $Q_0$ at the fixed accelerating field caused by the quench. The quench events considered are caused only by "hard" limiting factors (e.g., thermal breakdown), whereas multipacting or field-emission-related quenches are not considered in this study. The very low-compensated magnetic field ($<1$ mG) is achieved by adjusting the Helmholtz coils' current in order to cancel out the magnetic field naturally present in the vertical measurement cryostat ($\lesssim 5$ mG).

### A. Quenching in compensated ambient fields

The first series of quenches are performed in compensated external magnetic fields. All quenches are "hard," reached by increasing the rf field. As Fig. 2 clearly shows, no appreciable $Q_0$ degradation is observed after quenching tens of times in the compensated field (red star), meaning that no extra dissipation is introduced for all the investigated bare cavities even though they are prepared with different treatments. The same lack of degradation is also the case for a fully dressed nine-cell cavity treated with the LCLS-II nitrogen-doping recipe, for which the average magnetic field value achieved by compensation coils just before the quench is lower than 2 mOe.

This phenomenon is important, as it rules out all other possible mechanisms of magnetic flux generation and trapping during quench except for the static ambient field, as those will necessarily lead to a decrease in $Q_0$ even in zero ambient field. In other words, magnetic flux trapped at the quench spot is not generated intrinsically, but it is extrinsic to the cavity.

### B. Degradation in noncompensated ambient magnetic field and recovery by zero-field quenching

In the second series of experiments, a finite value of the magnetic field is applied outside of the cavity before quenching, and the degradation the $Q$ factor is clearly observed after a single and a number of quenches, as shown in Fig. 3(a), where $\Delta R_0(H)$ corresponds to the difference in averaged $R_0 = 270 \, \Omega/Q_0$ after and before any quenches. Then, in each case, the ambient field is again adjusted to as low as possible, and the cavity is quenched again several times [points with "0" field labels in Fig. 3(a)]. $Q_0$ can be totally recovered to its value just before any quenches, as is also observed in Ref. [8].

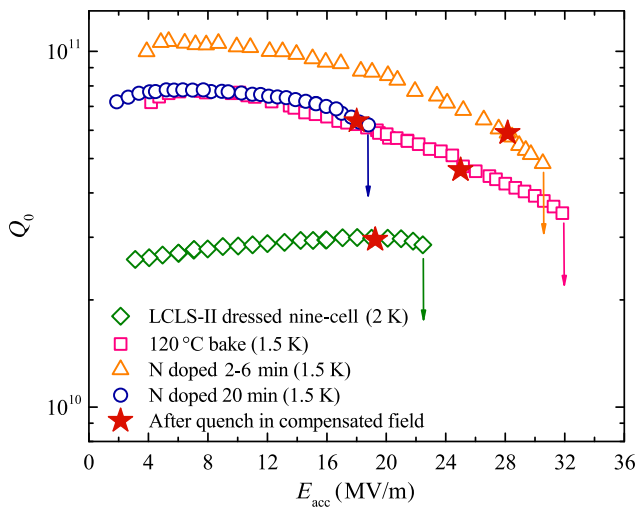

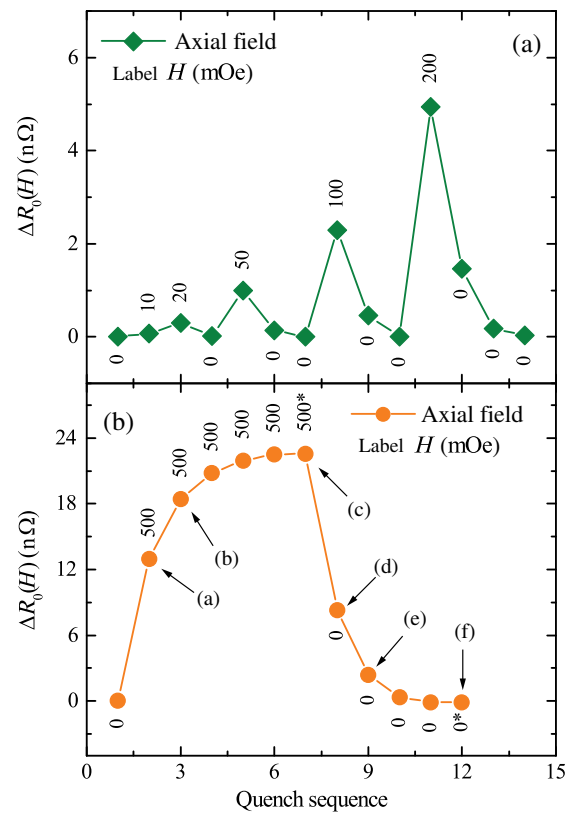

FIG. 2. $Q_0$ versus accelerating field curves acquired after a cooldown in compensated field before any quench. The red stars correspond to the $Q_0$ point acquired after quenching more than ten times in compensated external magnetic field.

FIG. 3. Quench study performed on cavity ACC002: (a) variation of the residual resistance due to quenches in the presence of external magnetic field; (b) saturation of the residual resistance due to multiple quenches in the same external field. The labels 0 indicate the condition of compensated field, while the symbol * refers to multiple quenches. $\Delta R_0(H)$ points that correspond to the $T$ maps of Fig. 4 are indicated with arrows and letters.

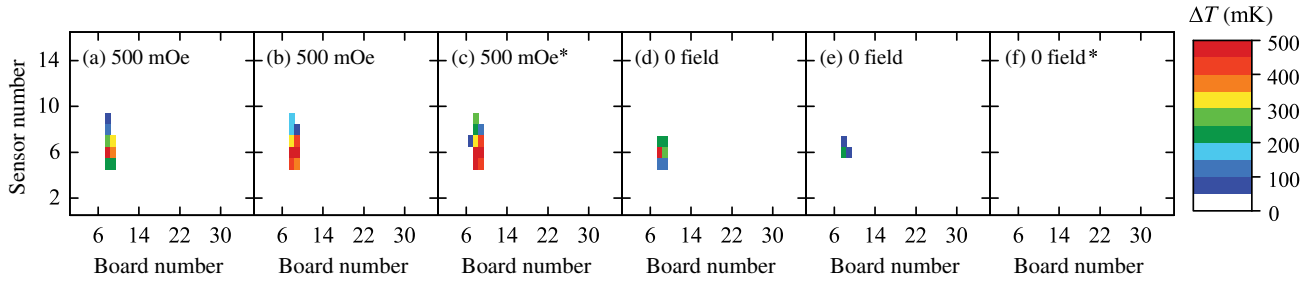

FIG. 4. Evolution of the dissipation due to trapped field at the quench spot for ACC002 after quenching: (a) a single, (b) two, (c) multiple times in 500 mOe, and after quenching (d) a single, (e) two, (f) multiple times in compensated field. The symbol star identifies multiple quenches. All the T maps were acquired at $E_{acc} = 18$ MV/m.

Interestingly, multiple quenches in the same field are needed for the residual resistance to reach a higher saturation value, as can be seen in Fig. 3(b). Such a saturation suggests that the maximum possible value of the magnetic flux trapped at the quench spot for a specific external magnetic field level is reached. Correspondingly, $Q_0$ can be totally recovered by several (and not a single) quench in the compensated low field.

It is possible to gain detailed insight into what happens during the multiquench saturation and recovery of $Q_0$ by analyzing the corresponding T maps. In Fig. 4, a sequence of temperature maps corresponding to the evolution of the magnetic flux trapped at the quench spot is shown. The corresponding residual resistance changes are highlighted with arrows in Fig. 3(b).

As it can be clearly observed, each of the quenches in 500 mOe [Figs. 4(a)–4(c)] leads to the progressive increase of the dissipation around the quench spot until the saturation is reached. Subsequent quenches in zero field [Figs. 4(d)–4(f)] cause the gradual decrease of the local dissipation until the prequench extra dissipation-free situation is attained, indicating the annihilation of the trapped flux.

The same $Q_0$ recovery effect is observed for all the cavities tested (Table I) independently on their surface preparation.

### C. Effect of the magnetic field orientation

The effect of the external field orientation on the $Q_0$ degradation during quench is studied on one of the cavities (AES011) in a series of single quenches in the presence of either nonzero axial or nonzero orthogonal components of the external field H with respect to the cavity beam axis. In Fig. 5, $\Delta R_0(H)$ is plotted as a function of the applied field. For the same magnetic field level, $\Delta R_0(H)$ is always higher for the orthogonal component, implying that most likely a larger amount of field is trapped.

In the Meissner state, the superconducting phase behaves as a perfect diamagnet, and the magnetic field is expelled from the cavity bulk and confined outside leading to the redistribution of the local field amplitudes at the cavity surface. In Fig. 6(a), a COMSOL Multiphysics simulation of the cavity in the Meissner state is shown for different field directions. As it is clearly seen, the final field configuration is strongly dependent upon the cavity axis orientation with respect to the applied field, and for a given axis orientation, the local field amplitude is strongly position dependent on the cavity surface as well.

The local field amplitude at the quench spot might be different than that measured by the fluxgate at the equator. Furthermore, for the same quench spot and for the same magnitude of the applied field, just varying the field orientation with respect to the cavity axis will change the local magnetic field value at the quench spot (as shown in Fig. 6). Therefore, different $Q_0$ degradation for different magnetic field components may be then just a manifestation of this purely geometrical effect.

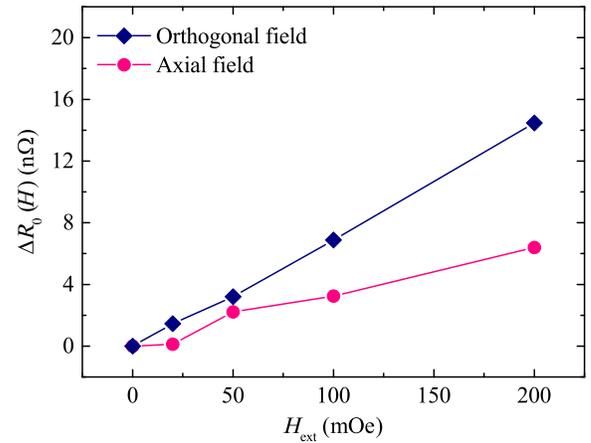

FIG. 5. Variation of the residual resistance of AES011 after single quenches for different values and orientations of the external magnetic field.

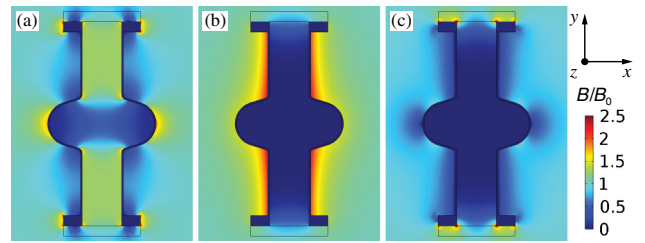

FIG. 6. Perfect Meissner effect simulation for different orientation of the magnetic field. (a) Field applied along y, (b) along z, and (c) along the x direction.

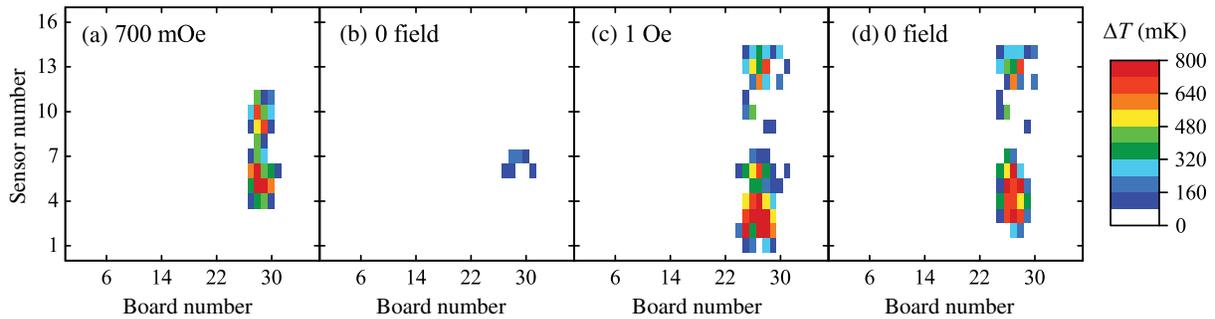

FIG. 7. $T$-map images acquired after the cavity AES019 is quenched in the presence of external magnetic field with the following sequence of magnitudes: (a) 700 mOe, (b) zero field, (c) 1 Oe, and (d) zero field. Such sequence shows the impossibility of $Q_0$ recovery after the cavity is quenched in 1 Oe. All the $T$ maps are acquired at $E_{\rm acc} = 17$ MV/m.

### D. Irrecoverable $Q_0$ degradation

Extending the experiments to higher values of magnetic field ($\gtrsim 1$ Oe), we find that once the cavity is quenched multiple times and the residual resistance is saturated, the cavity quality factor can be only partially recovered compared to its original value before the quench. It is demonstrated in Fig. 8 for AES019 where the residual resistance can be recovered to its original value only when the quench is performed in magnetic field values less or equal to 700 mOe. After the cavity is quenched several times in 1 Oe, its quality factor cannot be completely recovered anymore, even by quenching several times in the compensated field. The same behavior is also observed for cavities ACC002 and AES011, for which the magnetic field threshold above which the quality factor cannot be completely recovered is 700 and 300 mOe, respectively. Cavity AES014 is quenched in fields up to 700 mOe, but no irrecoverable $Q_0$ threshold is observed.

In Fig. 7, the evolution of the local dissipation due to the trapped field at the quench spot for the recoverable and irrecoverable $Q_0$ degradation of AES019 as registered by the $T$-map system is shown. The corresponding residual resistance variation is shown in Fig. 8. Figure 7(a) reveals the dissipation due to trapped magnetic flux after being quenched several times in 700 mOe. The ambient field is then compensated as much as possible, and the cavity is again quenched several times. Figure 7(b) indicates that most of the dissipation introduced by the previous quenches vanishes. Still, some flux remains trapped at the quench spot, probably because of the nonaxial field components of the field in the vertical test cryostat that cannot be compensated by the axial coils.

The external magnetic field is then set to 1 Oe, and the cavity is quenched several times again. The corresponding $T$ map in Fig. 7(c) shows that the fluxoid dissipation area has now spread out farther than after the 700-mOe quenches. In addition, the two dissipative "lobes" clearly emerge separated by a less-dissipative region in the middle. After the field is subsequently compensated and the cavity is quenched several times, no complete field annihilation occurs, while some redistribution of the magnetic flux is recorded, as shown in Fig. 7(d).

## IV. DISCUSSION

### A. Magnetic field redistribution during quench

To interpret the experimental results, it is first important to understand the dynamics of the magnetic flux during the quench event. In order to visualize such field dynamics, we use COMSOL Multiphysics to perform magnetic field simulations during a quench event.

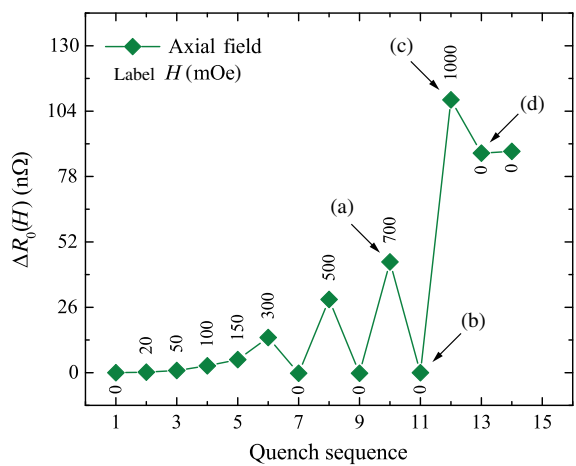

FIG. 8. Residual resistance evolution of AES019 after quenching in different field values. Every point in the graph corresponds to multiple quenches in the same applied field. The arrows indicate the data points that correspond to the $T$ maps of Fig. 7. The labels 0 indicate the condition of the compensated field.

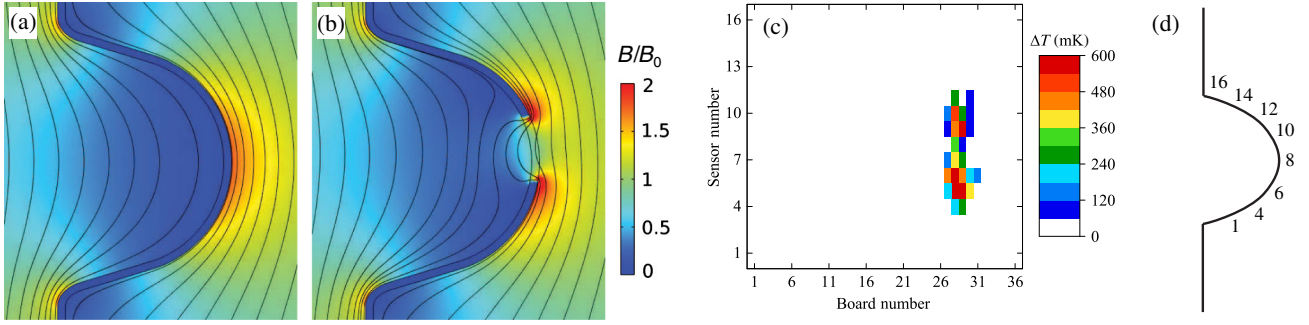

FIG. 9. Simulations of the magnetic field distribution around the quench spot: (a) before quench and (b) during quench. Color scale represents the ratio between the local magnetic field and the applied magnetic field. (c) $T$ map of AES019 after multiple quenches in 500 mOe (acquired at $E_{\mathrm{acc}} = 17$ MV/m); (d) schematics of the thermometer positions.

In the Meissner state before quench, the magnetic field (axial case is shown) is expelled from cavity walls and deflected around it, as shown in Fig. 9(a).

During the cavity quench, a normal-conducting hole opens on the cavity wall, which causes the redistribution of the magnetic field $H$. These changes can be described as driven by the magnetic force, which is given (per unit volume) by

$$\mathbf{f_m} = -\nabla\left(\frac{B^2}{2\mu_0}\right) + \frac{(\nabla \cdot \mathbf{B})\mathbf{B}}{\mu_0}, \quad (1)$$

where the first term corresponds to the magnetic pressure and the second one to the magnetic tension. The magnetic pressure is directed perpendicular to the magnetic field lines in the opposite direction to the field gradient. The magnetic tension is instead present only when the magnetic field is bent, and it has radial direction with aim directed toward the center of curvature. It introduces the same restoring action that the elastic force has when a stiff slab is bent. The magnetic tension then exerts a force to straighten out the bent magnetic field line.

Around the equatorial zone of the cavity, the magnetic field lines are denser and more bent in the Meissner state [Fig. 9(a)]; thus, both the magnetic pressure and the magnetic tension are directed towards the cavity wall. When the normal-conducting hole opens on the cavity wall, the magnetic field that is excluded from the cavity internal volume is now allowed to penetrate driven by the sum of the magnetic pressure and magnetic tension contributions.

In order to simulate the field configuration once trapped at the quench spot, we model the normal-superconducting boundary using the following approximate sigmoidal form of the field-dependent relative magnetic permeability $\mu_r(H)$ for a type I superconductor:

$$\mu_r(H) = \frac{1}{1 + e^{-(H-\alpha)/c}}, \quad (2)$$

where $\alpha = c \times \ln(0.0001) + H_{c2}$ is the parameter to ensure $\mu_r \cong 1$ for $H \geq H_{c2}$, $H_{c2} = H_0[1-(T/T_c)^2]$, $H_0 = 410$ mT is the second critical field at 0 K, $T_c = 9.25$ K is the critical temperature, and $c$ is the parameter that defines the slope of $\mu_r(H)$ through transition; the smaller $c$, the steeper the function.

To approximate best the sequential progression of the field redistribution during quench, we assume the pointlike heat source in the equatorial zone (but not right at the equator) where quench most frequently occurs and let the local temperature $T$ linearly increase with time. As the temperature increases, the local critical field $H_{c2}(T)$ of niobium decreases, and when the external applied field $H$ exceeds $H_{c2}(T)$ (so that $\mu_r \cong 1$), the external field starts penetrating the cavity wall at the heated region. Parameters used for niobium are thermal conductivity $\kappa = 30$ W m$^{-1}$ K$^{-1}$, the heat capacity $C_p = 0.126$ J kg$^{-1}$ K$^{-1}$, and the thermal boundary resistance at the niobium–liquid–helium interface $h_k = 5000$ W m$^{-2}$ K$^{-1}$.

Simulated field distribution for the moment in time when the normal zone opening is largest is shown in Fig. 9(b). It can be seen that the magnetic field lines form a semiloop inside the cavity volume, and when the quench region is cooled again below $T_c$, the field will be trapped in the form of fluxoids with the opposite directions of the magnetic field, entering on one side of the normal zone and exiting from another.

In Fig. 9(c), a $T$ map taken during rf measurements of cavity AES019 after a series of quenches in 500 mOe is shown, which is consistent with two lobes of dissipation corresponding to entry and exit points of the field lines in accordance with simulation [Fig. 9(d) can be used as a reference to understand the $T$-map image orientation and temperature-sensor locations].

### B. Quality factor recovery mechanism

The suspected mechanism at the basis of the $Q_0$ recovery phenomenon is the annihilation of the magnetic field trapped at the quench spot when the cavity is allowed to quench again but in a very low compensated field [15,16].

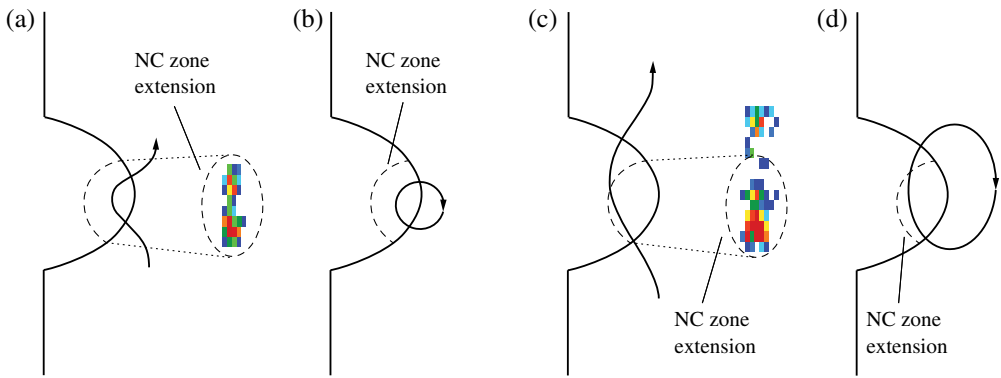

FIG. 10. Sketch of the magnetic field trapped at the quench spot: (a) after quench in the presence of external magnetic field (the T map shows a two-lobe-shaped dissipation pattern), (b) after the external field cancellation, (c) trapped magnetic flux after the flux migration (the T map shows two hot spots), and (d) field compensated after the magnetic flux migration.

With a finite magnitude of the applied magnetic field, the trapped magnetic field lines will create a closed loop passing through the two Helmholtz coils [Fig. 10(a)]. But when the external field is canceled out, to respect Ampère's law, the trapped magnetic field must be sustained only by the screening currents in the superconductor [Fig. 10(b)].

When the quench occurs, a normal-conducting region is created at the quench spot, and the trapped field vanishes upon the superconducting-normal-conducting transition annihilating the superconducting screening currents that sustained it. The dimension of the emerging normal-conducting region is governed by the total dissipated energy, which is set by the value of the maximum accelerating field at which the cavity quenches. Thus, if the quench field remains the same, the normal opening size should also be the same, and all the trapped field within the quench zone can be annihilated.

In this simple model, the trapped magnetic flux can be annihilated only if (i) it is trapped in the loop configuration, and (ii) if fluxoid entry and exit points are both inside the maximum extension of the normal-conducting area during the quench.

If the condition (ii) is not satisfied [e.g., Fig. 10(c)] and some of the fluxoid entry or exit points fall outside of the maximum extension of the normal-conducting zone during the quench, then even if the external field is compensated and the cavity is quenched again, those superconducting currents that sustain the trapped field outside of the normal opening will still exist, preventing the full annihilation of the trapped magnetic flux. Such a situation may occur either if the trapped flux is migrated away from the original location (discussed in detail below) or if the quench field is decreased due to the extra dissipation introduced by the trapped flux. The normal-conducting opening size will be then smaller than before and less field can be annihilated.

As we show above, it takes more than one quench to reach a saturation level of the increased surface resistance when quenching in nonzero field, and several quenches in zero field are typically needed to completely recover the quality factor, the more so the higher the trapped field. We interpret these findings as the possible manifestation of the finite time constant $\tau_B$ for the magnetic field configuration to change during the normal zone opening. The value of $\tau_B$ is determined by the damping time of eddy currents, which counteract the penetration of the magnetic field [17]. If the characteristic time of the normal zone existence $\tau_{NC} \sim 100$ ms [18] is shorter than $\tau_B$, then the magnetic field distribution in the quench area will only reach the steady state after a number of quenches $N \sim \tau_B/\tau_{NC}$.

Lastly, it is important to emphasize again that the quality factor recovery in the compensated field reveals the extrinsic nature of the magnetic flux trapped during the quench. If the magnetic flux is intrinsic to the cavity, or in other words, generated by the quench itself (i.e., local thermal currents), then the recovery of $Q_0$ can be achieved only by warming the cavity above $T_c$, as quenching in zero applied magnetic field will be only a source of extra dissipation as any other quench.

### C. Magnetic flux migration

In this subsection, we discuss a possible mechanism at the basis of the observed flux migration phenomenon.

The equation of motion of a single fluxoid in the absence of the drift current is described in Refs. [19,20]. In the present case though, fluxoids and antifluxoids are mutually connected sharing the same magnetic field lines. This means that the motion of the two lobes is coupled via the magnetic tension. The main action of such a force is to straighten the magnetic field lines that connect fluxoids and antifluxoids, pulling apart the two lobes. The modified equation of motion is then

$$\frac{\Sigma}{N}\frac{(\nabla \cdot \mathbf{B})\mathbf{B}}{\mu_0} - S \cdot \nabla T - \eta \mathbf{v} - f n_s e(\mathbf{v} \times \phi_0 \hat{\mathbf{u}}_n) - \mathbf{f_p} = 0. \quad (3)$$

The first term corresponds to the magnetic tension per fluxoid with **B** the trapped field, $\Sigma$ the normal surface area through which the magnetic field bent inside the cavity volume passes, and $N$ the number of fluxoids. The second term is the thermal force, whose direction depends only on the thermal gradient $\nabla T$, and which pushes a fluxoid toward colder regions [19–22]. $S$ is the transport entropy per unit length given by [19,20]

$$S = -\phi_0 \frac{\partial H_{c1}(T)}{\partial T} = 2\phi_0 H_0 \frac{T}{T_c}, \quad (4)$$

where $H_0$ is the lower critical field at 0 K (190 mT for clean niobium) and $\phi_0$ the flux quantum. The origin of the temperature gradient is the increased local rf dissipation at the trapped flux location, which makes the thermal force directed to spread the trapped flux around.

Here we consider trapped fields far below $H_{c2}$ and, therefore, neglect the interaction between different fluxoids, which is a rapidly decreasing function of the interfluxoid spacing. We also do not include the time-dependent Lorentz force acting between the surface screening currents in the cavity and the trapped flux, as its net effect is the oscillation of the surface segments of the magnetic flux around the stable equilibrium position [21,23,24].

The combination of the magnetic tension and thermal force will act against the Magnus force $f n_s e (\mathbf{v} \times \phi_0 \hat{\mathbf{u}}_\mathbf{n})$, the viscous damping drag force $\eta \mathbf{v}$, and the pinning force $\mathbf{f_p}$. Here, $n_s$ is the electron density, $f$ is the fraction of the Magnus force that is active, and the fluxoid motion viscosity per unit length $\eta$ is given by [25]

$$\eta = \frac{3}{2} \frac{\sigma_n \phi_0^2}{\pi^2 \xi_0 l}, \quad (5)$$

with $\sigma_n$ the normal electron conductivity, $\xi_0$ the coherence length, and $l$ the electron mean free path.

Fluxoids will start migrating when the sum of magnetic tension and thermal force becomes larger than the pinning force:

$$\frac{\Sigma}{N} \frac{(\nabla \cdot \mathbf{B}) \mathbf{B}}{\mu_0} - S \cdot \nabla T_k \geq \mathbf{f_p}. \quad (6)$$

Magnetic tension in Eq. (3) plays a crucial role, as it allows us to explain why the motion of lobes happens along a straight line. If the thermal force is the only driving force of flux migration, then the net motion will be isotropic; i.e., we will see the lobes becoming broader and more blurry as the local temperature is increased, driven by rf dissipation. What we see, instead, are the lobes moving in the opposite directions along the same line, as it is expected if the magnetic tension were also non-negligible. Both contributions are dependent on the amount of magnetic field trapped during the quench. The higher the trapped magnetic field, the larger the local thermal gradient appearing when the rf field is reestablished inside the resonator and, therefore, the larger the thermal force. Similarly, the magnetic tension term is proportional to $B^2$ and is higher for larger trapped fields.

Using the flux motion description, we can estimate the pinning potential strength that will correspond to the observed fluxoid migration thresholds. We approximate the pinning potential as an ideal inverse Lorentzian curve [26] that is acting on the whole flux line crossing the cavity wall,

$$U_p(x) = -\frac{U_0}{1 + [(x - x_0)/\xi]^2}, \quad (7)$$

where $x_0$ is the initial fluxoid position, $U_0$ is the pinning potential depth, and $\xi = (1/l + 1/\xi_0)^{-1}$ is the effective coherence length that determines the depth of the pinning potential well.

The equation of motion [Eq. (3)] is then numerically solved with the program Mathematica, considering the flux line length equal to the cavity wall thickness and small displacements near the pinning potential minimum, which allows us to take the radius of curvature for the magnetic field lines constant. The number of fluxoids $N$ is estimated as equal to the lobe area $A$ times the trapped magnetic field $B$, over the flux quantum: $N = AB/\phi_0$. Fixed parameters for the simulation are trapped field $B = 1$ Oe, thermal gradient $|\Delta T| \approx 1.7$ K cm$^{-1}$, and the flux area $A = 4\pi$ cm$^2$ [both estimated from the $T$ map in Fig. 7(d)], the coherence length $\xi_0 = 39$ nm, and the mean free path $l = 100$ nm. Simulation starts from $t = 0$ when the normal-conducting zone just finishes closing, and the trapped flux starts moving driven by thermal and magnetic forces.

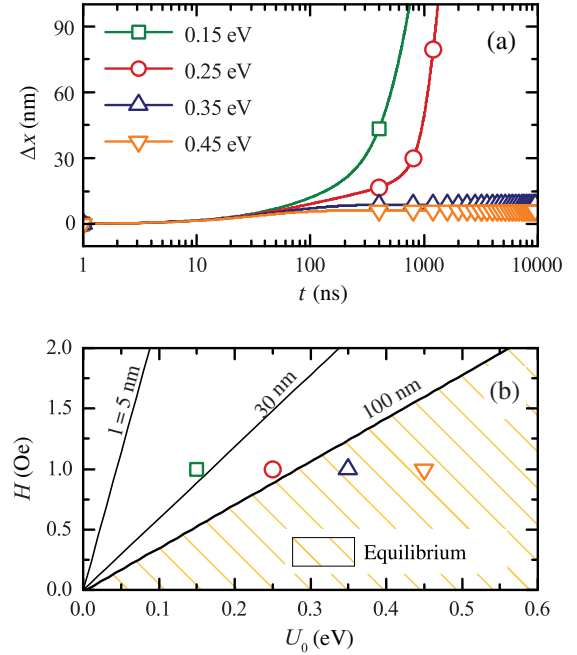

FIG. 11. The numerical solutions of the fluxoid motion equation [Eq. (3)] for the displacement $\Delta x$ as a function of time for different pinning potentials is plotted in (a). The magnetic field considered in the calculation is 1 Oe, with mean free path 100 nm. The separatrix line between equilibrium and fluxoid migration is plotted in (b) [see Eq. (6)]. Different lines correspond to different mean free paths. The points refer to magnetic field $B$ and potential $U_0$ chosen for the numerical solutions in (a).

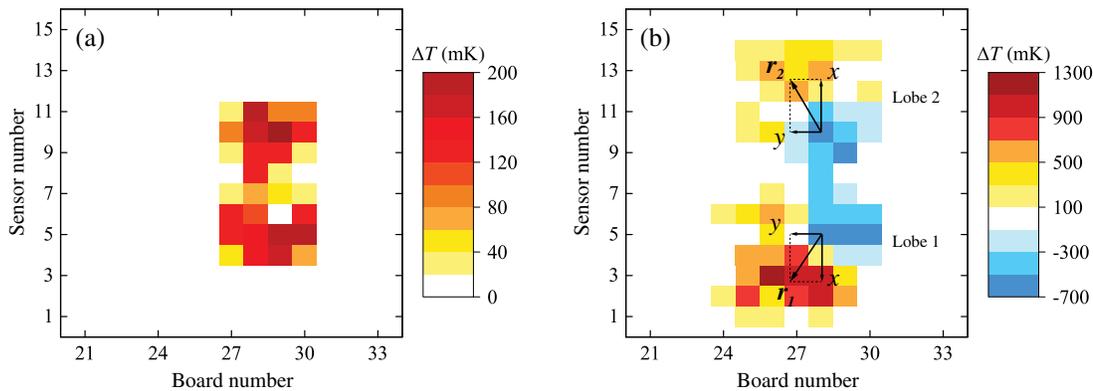

FIG. 12. Redistribution of the magnetic flux. In (a) the difference between the $T$ map acquired after quenching in 700 and in 500 mOe is reported, while in (b) between quenching in 1 and in 700 mOe. The flux redistributes or adds from negative to positive value regions.

Figure 11(a) shows the simulation results for the fluxoid displacement $\Delta x$ as a function of time for different values of the pinning potential $U_0$. In the case where $U_0$ is such that Eq. (6) is not satisfied, the fluxoid displacement approaches an equilibrium value of $\Delta x \approx 5$–10 nm from the potential minimum after approximately 1 $\mu$s.

When, instead, Eq. (6) is respected, the fluxoid motion deviates from the previous situation. Initially, the fluxoid drifts, experiencing the pinning force, but after a certain time (depinning time), it depins, and its displacement from the pinning site grows drastically [as shown in Fig. 11(a) for $U_0 = 0.15$ and 0.25 eV]. It is important to notice that deeper pinning potential wells correspond to longer depinning times.

Figure 11(b) shows the separatrix line between the regions of constrained pinned and unconstrained depinning [Eq. (6)] motion of flux for pairs of trapped magnetic field $B$ and pinning potential $U_0$ values and different mean free path values (labels on the lines). For the points from Fig. 11(a), the separatrix falls in between the points for $U_0 = 0.25$ and 0.35 eV, giving the threshold value of $U_0 \simeq 0.28$ eV, which is a reasonable value for pure metals [26].

This simple migration model can also explain why different thresholds are observed. Assuming that different cavities might have different pinning potentials, we expect that the magnetic field threshold will be affected from that—the higher the pinning potential, the higher the magnetic field threshold for the migration phenomenon.

An insightful analysis is shown in Fig. 12 for the redistribution of the trapped flux for different quenches for cavity AES019. The relative changes are obtained by subtracting $T$ maps acquired after quenches in different external field values. Negative $\Delta T$ values in Fig. 12 reveal zones from which the flux moves away, positive $\Delta T$ zones indicate the areas where the flux is added, and $\Delta T = 0$ corresponds to areas where the flux remains the same. In Fig. 12(a), such an incremental difference is shown for the trapped flux dissipation after quenching in 700 mOe, as compared to quenching in 500 mOe. Clearly, no field redistribution occurs, and simply more flux is trapped. This case is consistent with the analysis above for the trapped field that is not high enough, and fluxoids are displaced by only tens of nanometers from their initial postquench position; therefore, no migration is observed with the $T$ map.

The pattern is clearly different in Fig. 12(b), which shows the difference after quenching in 1000 mOe with respect to 700 mOe. In this case, a clear redistribution of the trapped magnetic field is found. The flux trapped after quenching in 1000 mOe migrates away from the central zone: lobe 1 follows the path indicated by the unit vector $\mathbf{r_1}$, while lobe 2 follows along the direction of unit vector $\mathbf{r_2}$. Such a situation can emerge when Eq. (6) is satisfied and the pinning force is overwhelmed by magnetic tension and thermal forces, and, consequently, two lobes are pushed farther apart.

Both $\mathbf{r_1}$ and $\mathbf{r_2}$ vectors can be decomposed into their components $x$ and $y$, and, surprisingly, both possess a nonzero component along $y$ [see Fig. 12(b)]. As the $y$ component is orthogonal to the path dictated only by the magnetic tension—along $x$—it should be attributable to some force orthogonal to the fluxoid motion—Magnus force being the likely suspect. The direction of the orthogonal motion is the same for the two lobes, which is also compatible with the Magnus force acting on outgoing flux from lobe 1 and incoming flux from lobe 2. For extremely pure superconductors, the Magnus force is shown to play an important role [19,20], and high-purity Nb (residual resistivity ratio approximately 300), out of which cavities are made for our studies, may also fall into this class.

## V. CONCLUSIONS

In this paper, we demonstrate that the origin of the magnetic flux trapped at the quench spot is purely extrinsic to the superconducting cavity. The key supporting finding is no extra dissipation ($Q_0$ is not affected) introduced by quenches in zero magnetic field, allowing every intrinsic mechanism of flux trapping, i.e., generation of thermal currents or trapping of rf field, to be ruled out. Our conclusion is further corroborated by (i) the clear relation of dissipation introduced by quenching to the orientation of the applied magnetic field and (ii) the possibility to totally recover the quality factor—without warming the cavity above $T_c$—by compensating the external field and requenching several times. Interestingly, for larger values of the ambient field (that allows magnetic and thermal forces larger than the pinning force), the $Q$-factor recovery may become impossible due to the outward flux migration beyond the normal zone opening during quench.

We attribute the flux migration process to the synergistic action of the thermal force generated by thermal gradients caused by the rf dissipation of the trapped flux itself and the magnetic tension on the trapped flux lobes against the pinning force. If one of the two lobes migrates outside of the maximum extension of the normal-conducting region during the quench, the $Q_0$ recovery is not possible. A transverse component in the fluxoids' migration path is also observed, which is compatible with the presence of a non-negligible fraction of the Magnus force.

Different magnetic field thresholds for the migration are observed. Based on simulations, this phenomenon is most likely introduced by different values of the pinning potential that affects the migration—the deeper the pinning potential, the larger the minimum value of trapped magnetic field needed to ignite the flux migration.

## ACKNOWLEDGMENTS

Fermilab is operated by Fermi Research Alliance, LLC under Contract No. DE-AC02-07CH11359 with the United States Department of Energy. The work is supported by the DOE HEP Early Career Grant of A. Grassellino, and DOE NP Early Career Grant of A. Romanenko.